\documentstyle[namedreferences,psfig]{kluwer}
\newcommand{\citeN}[1]{\citeauthor{#1}\ (\citeyear{#1})}
\newcommand{\citeNP}[1]{\citeauthor{#1},\ \citeyear{#1}}

\newcommand{\apj}{{\it Astrophys.~J.}}

\newcommand{\aap}{{\it Astron.~Astrophys.}}

\newcommand{\solphys}{{\it Solar~Phys.}}

\newcommand{\ujlisti}{
\itemsep=0 em
\parsep=0.5 em
\partopsep=0.25 em
\topsep=0 em}
\newcommand{\ujlistii}{
\itemsep=0 em
\parsep=0.5 em
\partopsep=0.25 em
\topsep=0 cm}

\newenvironment{lista}{\begin{list}{--}{\ujlisti}}{\end{list}}

\renewcommand{\[}{\begin{equation}}
\renewcommand{\]}{\end{equation}}
\newcommand{\scri}{\scriptsize}

\newcommand{\ov}{\overline}

\def\rvar{\tilde r}

\def\ga{\mathrel{\mathchoice {\vcenter{\offinterlineskip\halign{\hfil
 $\displaystyle##$\hfil\cr>\cr\sim\cr}}}
 {\vcenter{\offinterlineskip\halign{\hfil$\textstyle##$\hfil\cr
 >\cr\sim\cr}}}
 {\vcenter{\offinterlineskip\halign{\hfil$\scriptstyle##$\hfil\cr
 >\cr\sim\cr}}}
 {\vcenter{\offinterlineskip\halign{\hfil$\scriptscriptstyle##$\hfil\cr
 >\cr\sim\cr}}}}}

\newcommand{\bext}{b_{\infty}}
\newcommand{\alphanu}{\alpha_\nu}
\newcommand{\ti}{t_i}

\begin{document}                                                                                   
\begin{article}
\begin{opening}

\runningtitle{Making Sense of Sunspot Decay, II}
\runningauthor{Petrovay, {Mart\'{\i}nez~Pillet} \& van~Driel-Gesztelyi}

\title{MAKING SENSE OF SUNSPOT DECAY}
\subtitle{II: Deviations from the Mean Law and Plage Effects}                                         

\author{K. \surname{Petrovay}}
\institute{Instituto de {Astrof\'\i sica} de Canarias, 
	   La Laguna, Tenerife, E-38200 Spain, and\\
	   E\"otv\"os University, Dept.~of Astronomy, Budapest, Pf.~32,
	   H-1518 Hungary} 
\author{V. \surname{Mart\'{\i}nez~Pillet}}
\institute{Instituto de {Astrof\'\i sica} de Canarias, 
	   La Laguna, Tenerife, E-38200 Spain} 
\author{L. \surname{van~Driel-Gesztelyi} } %\thanks{%
           % Affiliated to WKAP, Dordrecht.}}
\institute{Observatoire de Paris, 
	   DASOP, F-92195 Meudon Cedex, France, and\\
	   Konkoly Observatory, Budapest, Pf. 67, H-1525, Hungary}

\date{[{\it Solar Physics}, in press  (1999)]}

\begin{abstract} 
In a statistical analysis of Debrecen Photoheliographic Results sunspot area
data we find that the  logarithmic deviation $(\log D)'$ of the area decay rate
$D$ from the parabolic mean decay law (derived in the first paper in this
series) follows a Gaussian probability distribution. As a consequence, the
actual decay rate $D$ and the  time-averaged decay rate $\ov D$ are also
characterized by approximately lognormal distributions, as found in an earlier
work. The correlation time of $(\log D)'$ is about 3 days. We find a
significant physical anticorrelation between $(\log D)'$ and the amount of
plage magnetic flux of the same polarity in an annulus around the spot on Kitt
Peak magnetograms. The anticorrelation is interpreted in terms of a
generalization of the turbulent erosion model of sunspot decay to the case when
the flux tube is embedded in a preexisting homogeneous ``plage'' field. The 
decay rate is found to depend inversely on the value of this plage field, the
relation being very close to logarithmic, i.e.\ the plage field acts as
multiplicative noise in the decay process. A Gaussian probability distribution
of the field strength in the surrounding plage will then naturally lead to a
lognormal distribution of the decay rates, as observed. It is thus suggested
that, beside other multiplicative noise sources, the environmental effect of
surrounding plage fields is a major factor in the origin of lognormally
distributed large random deviations from the mean law in the sunspot decay
rates.
\end{abstract}
\end{opening}

%\motto{This is a motto\\
%      For details see section~\ref{motto}}
  
%\keywords{LaTeX, Kluwer Style File, Authors' Instructions}

%\abbreviations{CM -- central meridian; DPR -- Debrecen Photoheliographic 
%	       Results; GPR -- Greenwich 
%	       Photoheliographic Results; MSH -- millionth solar hemisphere; 
%	       MSHER -- MSH equivalent radius}

%\classification{JEL codes}{D24, L60, 047}

\section{Introduction}
%\footnotemark{}\footnotetext{{\it Solar Physics}, in press  (1999)}
The very large and apparently random individual differences in the area decay
rates of sunspots have constituted a serious obstacle in the way of an empirical
determination of a mean decay law. With time it has been recognized that the
study of the {\it distribution\/} of decay rates itself may be of key importance
for the proper statistical analysis of the decay, as most of the widely applied
tools of mathematical statistics (e.g.\ least-square fits) are based on the tacit 
assumption of one particular (usually the Gaussian) probability distribution.
The realization that the mean area decay rates $\ov D$ of sunspots are
distributed lognormally (\citeNP{VMP+:periph.decay}, hereafter MMV93) was
therefore an important step towards the proper understanding of the statistical
regularities underlying the apparently haphazard sunspot decay process. A
lognormal distribution implies that it is the {\it logarithm\/} of the decay
rate $D$ that is normally distributed, and thus it is $\log D$ that should be
used as the dependent variable in least-square fits and other standard
statistical procedures. This realization has helped us in the first paper of this
series (\citeNP{Petrovay+vDG:decay1}, hereafter Paper I; see also
\citeNP{Petrovay:supdif}) to finally determine, 
by rigorous statistical methods and at a quite convincing confidence level, the
mean law governing sunspot decay. It was found that an ``idealized'' sunspot 
following this mean law exactly, with no random deviations, would decay
according to the law 
\[ D_{\mbox{\scri ideal}}=C_D \,r/r_0 \qquad C_D=32.0\pm 0.26 .     \label{eq:declaw} \]
where $D_{\mbox{\scri ideal}}$ is the area decay rate $D$ for an idealized spot
in units of MSH/day (MSH: one millionth solar hemisphere), $r$ is the {\it
equivalent radius\/} of the spot (i.e.\ the radius of the circle with the same
area as that of the spot), while $r_0$ is its maximal equivalent radius. (The
corresponding areas will be denoted by $A$ and $A_0$, respectively.) It is
clear that as $D=-\dot A$ and $r\sim A^{1/2}$, Equation (\ref{eq:declaw})
implies a {\it parabolic decrease} of the spot  area with time, in contrast to
the previously widely held view that the decay  is linear.

For real spots the actual daily decay rate $D$ shows random deviations from 
the law  (\ref{eq:declaw}). It follows from the considerations of the first
paragraph above that in the derivation of the law (\ref{eq:declaw}) it is the
{\it logarithm} of $D$ that had to be averaged, i.e.\ the mean decay law for
real sunspots is 
\[ \langle \log D\rangle =\log (C_D \,r/r_0)  \label{eq:precdeclaw} \]
(Here we are using the notation $\langle a\rangle$ for the ensemble average
over many spots at the same phase of decay $r/r_0$, as opposed to the time
average over the decay phase of one spot, $\ov a$. Deviations from the ensemble
average are denoted as $a'=a-\langle a\rangle$.) 

Having determined the mean decay law, in the present paper we turn our attention
to the {\it deviations\/} from this mean law. There are several important issues
to clarify in this respect.
\begin{lista}
\item The analysis of Paper I assumed that $(\log D)'$ follows a Gaussian
distribution. In the case of a parabolic  decay law like (\ref{eq:declaw})
however it is not possible to have a {\it strictly\/} normal distribution
simultaneously for $(\log D)'$, {\it and\/} also for $\log\ov D$ and $\log D$, 
as suggested in MMV93. In Section~2 we therefore consider the question, which
one of the proposed lognormal distributions is the fundamental and exact one,
and which are those that are only approximate and appear as secondary
consequences of the fundamental distribution.
\item The decay law (\ref{eq:declaw}) implies that the decay process has a {\it
``memory''\/}: in order to ``know'' the ``right'' decay rate, the spot must
remember its maximal radius $r_0$ at all later phases of its decay. The
question arises, how can this long-term memory be reconciled with the presence
of random  fluctuations in the decay rate? In Section~3 we first estimate the
correlation time of the fluctuations, and then, considering the possible
explanations, we conclude that the ``memory'' responsible for both the
long-term systematics and for the short-term fluctuations can be identified
with the amount of flux stored in the plage around the spot. This implies that
decay rate fluctuations should be correlated with the amount of this flux.
\item In order to check the above prediction, in Section~4 we compare the
amount of magnetic flux in an annulus surrounding the spot, as determined from
Kitt Peak magnetograms, with the decay rate. We indeed find a significant
anticorrelation between the two quantities. Interpreting this empirical
anticorrelation, in  Section~5 we generalize the turbulent erosion model to
account for a plage field.
\end{lista}
Finally, in the Conclusion we discuss and summarize the main factors leading to,
and the physical interpretation of the fluctuation in the decay rates.

Throughout the paper, we use the same data set as in Paper I, taken from the
Debrecen Photoheliographic Results (DPR) 1977--78 (\citeNP{DPR},
\citeyear{DPR78}). This consisted of 3990 area measurements of 476 different 
spots. For a detailed discussion of the data selection and reduction
process we refer the reader to Section~3 of Paper I. In order to extend the
data set, we made an attempt to include data from the more extensive Mt Wilson
catalogue. However, owing to the lower precision of those data and the problems
with the day-to-day identification of sunspots we decided to use the DPR data
only for the present analysis. Nevertheless, the comparison of the two catalogues
offered some interesting conclusions, summarized in Appendix A.

\section{The distribution of decay rate fluctuations}
\begin{figure}
%\vspace{1 cm}  % Amount of vertical space needed
\centerline{\psfig{figure=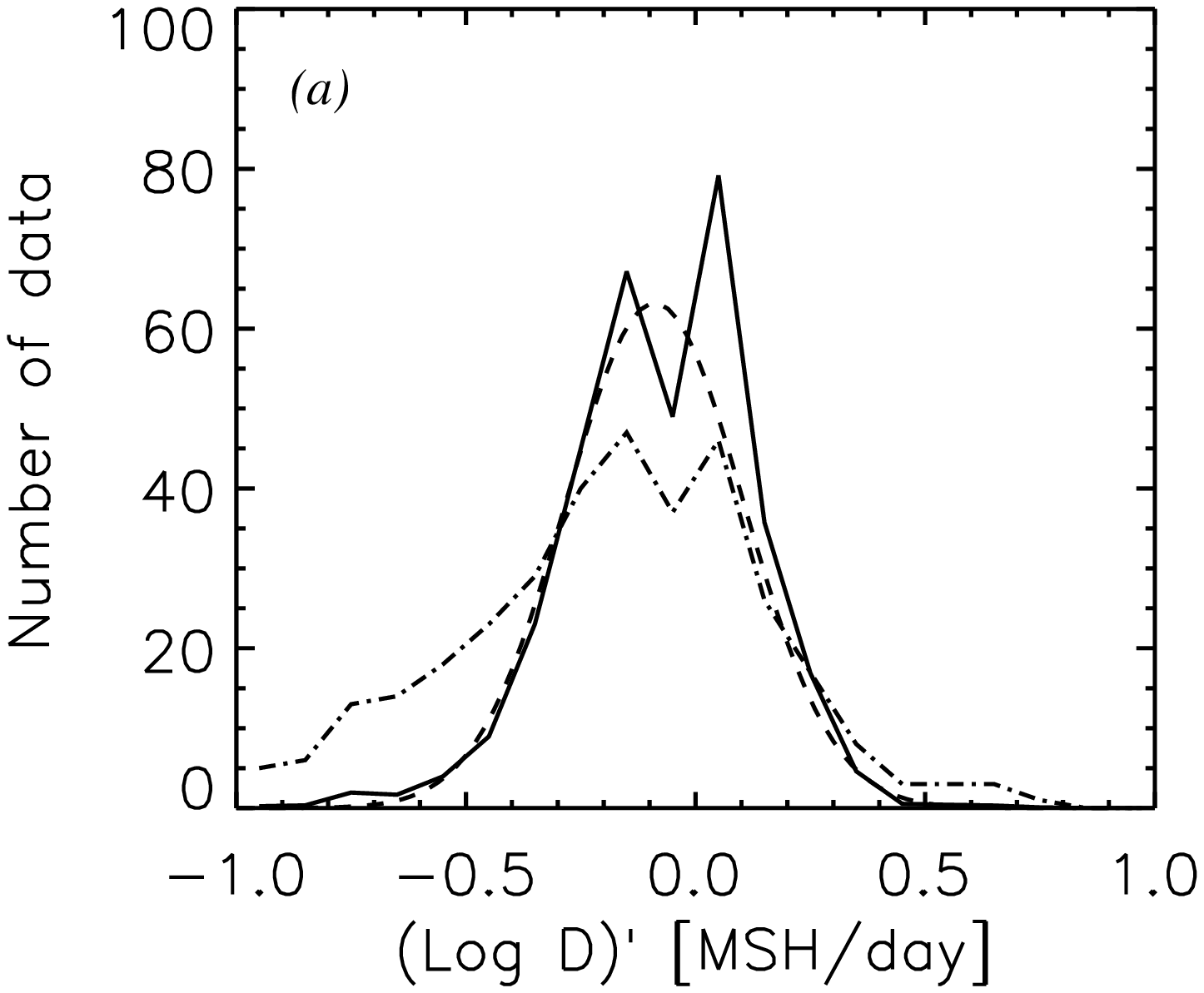,width=7.0 cm}\hskip -0.6 cm
            \psfig{figure=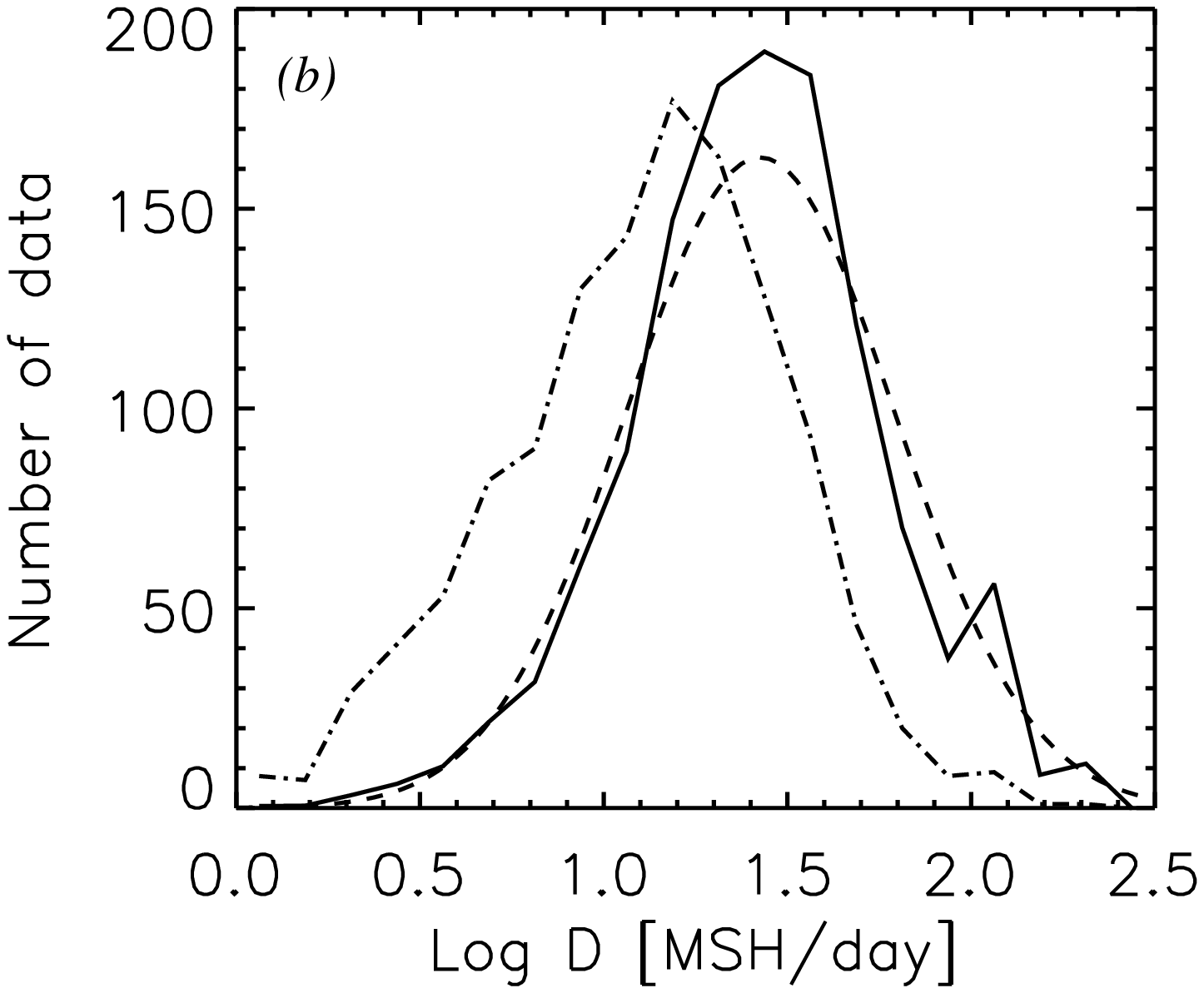,width=7.0 cm}}
\caption{Histograms of {\it (a)\/} $(\log D)'$ and {\it (b)\/} $\log D$ for all
spots that died on the visible hemisphere. (Fig.~1a contains data for
$r/r_0>0.7$ only.) Dash-dotted: raw histograms; solid: histograms corrected for
selection effect; dashed: Gaussian fits. The parameters of the fits are 
$\langle{(\log D)'}\rangle=-0.09$, $\sigma=0.19$ for {\it (a)\/} and 
$\langle{\log D}\rangle=1.25$, $\sigma=0.23$ for {\it (b)}.}
\label{fig:flucdist}
\end{figure}

Figure~\ref{fig:flucdist}a shows the histogram of $(\log D)'$, as defined after 
Equation  (\ref{eq:precdeclaw}), for our sample (dash-dotted line). This
original histogram  is distorted by the selection effect described in
Section~3.3.2 of Paper I: this effect essentially consists of a lower expected 
number of observations for spots with higher decay rates, and thus shorter 
lifetimes. (The histogram contains data points with $r/r_0>0.7$ only, to avoid
the regime most affected by this bias.) Correcting for the bias using the
method described there, we obtain the histogram shown by the solid line. A
Gaussian fit to this histogram (dashed) is found to be a fairly good
representation of the data.  ($Q=\Gamma[(N-1)/2,\chi^2/2]=0.02$; the skewness
and kurtosis are $-0.20\pm 0.21$ and $0.67\pm 0.52$, respectively, i.e.\ not
significantly different from 0.) This normal distribution of $(\log D)'$
confirms the correctness of the analysis of Paper I, based on least-square fits
with $\log D$ as the dependent variable.

\begin{figure}
%\vspace{1 cm}  % Amount of vertical space needed
\centerline{\psfig{figure=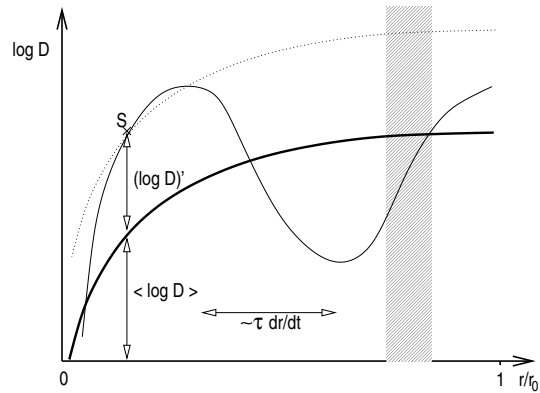,width=7.0 cm}}
\caption{Sketch for the explanation of the relation between the lognormal
distributions of different decay rates. $S$ is a generic data point in the
plane $r/r_0$--$\log D$ where the thick curve represents the mean decay law
(1). The thin curve is a decay curve for a generic sunspot with a random 
logarithmic  deviation $(\log D)'$ varying on a timescale $\tau$, while the
dotted curve is the decay curve of a hypothetical suspot with a constant
logarithmic deviation from the mean law, $(\log D)'=$const.}
\label{fig:figref}
\end{figure}

In MMV93 it was found that, within the errors, $\log\ov D$ is normally
distributed. Furthermore, based on a statistics of Mt Wilson data
(\citeNP{Howard:decay}) it was suggested that $\log D$ should also show a
Gaussian distribution. However, our above finding on the normal distribution
of $(\log D)'$, together with the parabolic  mean decay law
(\ref{eq:precdeclaw}), implies that the distributions studied in MMV93 cannot
be {\it exactly\/} Gaussian. 

To see this first for $\log D\equiv\langle\log D\rangle +(\log D)'$, let us
consider the plane $r/r_0$--$\log D$ (Figure~\ref{fig:figref}). The normal
distribution of $(\log D)'$ with zero mean implies that, for a narrow range of 
$r/r_0$ (shaded area), $\log D$ is also normally distributed around  a mean
value $\log(C_D r/r_0)$. This mean depends on $r/r_0$; hence, the  distribution
of $\log D$ in the complete sample (essentially, the distribution of all data
points in the plane projected on the ordinate) will deviate from a Gaussian.
(This is in contrast to the case of linear decay where the mean decay law,
indicated by the thick curve on Figure~\ref{fig:figref}, would be a horizontal
line, independent of $r/r_0$.) Indeed, in  Figure~\ref{fig:flucdist}b we
present the histogram of $\log D$: while at first glance the quality of the fit
does not seem to be very different from the previous case, a quantitative
analysis shows that the Gaussian fit is insufficient to explain the data.
($Q<10^{-5}$; the skewness and kurtosis are $-0.77\pm 0.13$ and $-0.13\pm
0.33$, respectively, i.e.\  the skewness significantly differs from 0.) The
deviation from gaussianity is relatively mild owing to the fact that the number
of data points is higher for higher values of $r/r_0$ (cf.\ Fig.~1 of Paper I)
where the thick curve representing the mean law  is not far from horizontal.
The distribution of $\log D$ is therefore indeed {\it approximately\/}
lognormal, as proposed in MMV93; but, as we pointed out, the deviation is
statistically significant. 

Let us now turn our attention to the time-averaged decay rate, defined as 
\[ \ov D=\frac 1{t_2-t_1} \int_{t_1}^{t_2} D \,dt  .
\]
Defining $\log C_D'\equiv\log C_D +(\log D)'$, according to Equation (2) we may 
formally write
\[  D= C_D' r/r_0 .  \label{eq:formaldeclaw}
\]
For a hypotetical sunspot with $(\log D)'=$const.\ (dotted curve in 
Figure~\ref{fig:figref}) $C_D'=$const., so from Equation 
(\ref{eq:formaldeclaw})
\[ \log \ov D = \log C_D'+\log \ov{r/r_0}   \label{eq:ovD} 
\]
where the last term is a constant for all spots, as for a parabolic decay law,
$r$ decreases linearly with time. ($\ov{r/r_0}=0.5$ if $t_1$ and $t_2$
correspond to the area maximum and the total disappearance of the spot,
respectively.) So for such spots if $(\log D)'$ is normally distributed, so is
$\log C_D'$ and $\log\ov D$.  For real sunspots, however, $(\log D)'$=const.\
does not apply; instead, their actual decay rate will fluctuate around the mean
law with some timescale $\tau$ (thin curve in Figure~\ref{fig:figref}), and the
above reasoning does not hold. Note that if the mean decay law were
linear ($\langle \log D\rangle=$const.), the resulting distribution of $\log
\ov D$ would still be normal, albeit with a lower dispersion $\sigma$, as in
that case the last term in Equation (\ref{eq:ovD}) would not be present.

Thus, $\ov D$ in general cannot be {\it exactly\/} lognormally distributed; but
its actual distribution may still appear to be close to lognormal provided
either that $(\log D)'$ does not change very much in the time interval
$[t_1,t_2]$ (i.e.\ that $t_2-t_1<\tau$) for the majority of spots in the sample, 
{\it or\/} that $r/r_0$ does not change very much in the same time interval so
that the mean decay rate may be considered nearly constant in that period of
time. 

For a time average taken over  the whole decay phase, $t_1$ should be the time
of area maximum, and $t_2$ the end of the life of the spot. The number of spots
in our sample for which both their birth and death were observed is, however,
insufficient for a precise empirical determination of the distribution of
$\log\ov D$. 

One way to increase the number of spots in the sample is to define $t_1$ and
$t_2$ as simply the first and last observations of a given spot, without
respect to whether these coincide with the birth/death of the spot. However, we
have just seen that for the departure from a Gaussian distribution to be
significant it is important that we have $t_2-t_1 > \tau$ where $\tau$ is the
correlation time of decay rate fluctuations {\it and\/} that $r/r_0$ take a
sufficiently wide range of values during the observation period. As these
conditions are only valid for a relatively low fraction of all spots, the
overall distribution derived in this manner may be expected to  stay  quite
close to Gaussian. This may explain the findings of MMV93, where the method
applied to determine $\ov D$ was indeed crudely the same as outlined in this
paragraph.

We therefore conclude that the probability distribution of $(\log D)'$ is close
to normal, and we propose that it is this distribution that should be
considered as fundamental both in a physical and in a mathematical sense. The
approximately (but not exactly) Gaussian character of the distributions  of
$\log D$ and $\log\ov D$ is then a secondary consequence of that fundamental
distribution. 

\begin{figure}
%\vspace{1 cm}  % Amount of vertical space needed
\centerline{\psfig{figure=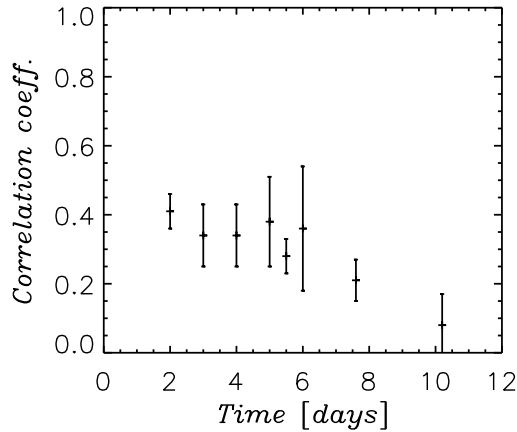,width=8 cm}}
\caption{Autocorrelation of $(\log D)'$ for sets of data pairs of
different mean time separation.}
\label{fig:autocorr}
\end{figure}

\section{The memories of a sunspot}
Figure~\ref{fig:autocorr} presents the autocorrelation of $(\log D)'$ for sets
of data pairs of different mean time separation (data pairs separated by 
1,2,3,$\dots$ consecutive observations) . It is apparent that the
autocorrelation time (where the autocorrelation sinks below $1/e$) is crudely
$\sim 3^{\mbox{\scri d}}$, to be compared with a typical spot lifetime of $\sim
7^{\mbox{\scri d}}$ in our sample. (This latter typical lifetime is entirely
due  to selection effects, but this is irrelevant for the present argument.)

The existence of random time-dependent fluctuations in the decay rate with a
correlation time shorter than the typical lifetime of spots leads to an
apparent conceptual problem.  As already mentioned in the Introduction, the
mean decay law (\ref{eq:declaw}) has a curious property: it implies that
sunspots have some ``long-term memory'' of their maximal radius even towards
the end of their lives. Two small sunspots of identical size will have
different expected daily decay rates if their maximal sizes were different. But
if the decay process is subject to a random ``noise'' with a short correlation
time, how is it possible for it to still possess a long-term ``memory'' in a
statistical sense?

The turbulent erosion model of sunspot decay (\citeNP{Petrovay+FMI:erosion},
hereafter PM97) successfully predicted the law (\ref{eq:declaw}). In this model
a ``memory'' effect is present due to the magnetic flux lost from the flux tube 
and piled up in its surroundings. The amount of this flux will influence the
steepness of the field gradient just outside the current sheet (CS), which in
turn determines the propagation speed of the CS (cf.\ Equation (10) in PM97).
Thus, in the framework of this model any fluctuations in the decay rate must be
due to fluctuations in the amount of the flux outside the tube. This flux can
be considered the sum of two contributions:
\[ \Phi=\Phi_{\mbox{\scri lost}} + \Phi_{\mbox{\scri plage}} 
    \label{eq:fluxterms} \]
where $\Phi_{\mbox{\scri lost}}$ is the flux lost from the spot, while 
$\Phi_{\mbox{\scri plage}}$ is the flux of independent origin. This latter
contribution is partly due to plage fields $\Phi_{\mbox{\scri pl,0}}$ which 
are originally present in the area, and partly to plage fields
$\Phi_{\mbox{\scri pl}}'$ which are carried  into/out of the neighbourhood of
the  flux tube by horizontal flows, flux loss from other spots, and so on:
\[ \Phi_{\mbox{\scri plage}}=\Phi_{\mbox{\scri pl,0}}+\Phi_{\mbox{\scri pl}}'
\]
 Both terms in Equation (\ref{eq:fluxterms}) show random fluctuations in time.
Now averaging the  equation we obviously have a relation for the expected value
of the flux around the spot (and therefore for the expected value of the decay
rate) as a functional of the mean decay law (first term on the r.h.s.\ of 
Equation~(\ref{eq:fluxterms})) and as a
function of  the plage  flux $\Phi_{\mbox{\scri pl,0}}$ originally present in
the site of the spot  (second term; the other contributions $\Phi_{\mbox{\scri
pl}}'$ to this term can be supposed to have zero mean).   Specifically, if 
$\Phi_{\mbox{\scri plage}}=0$, we are back to the models presented in PM97; the
more general case will be treated in Section~5 below. The initial conditions
thus fully determine the expected value of $D$ at any later time. It is thus
possible to have a long-term``memory'' in the statistical sense while
short-term fluctuations are still present.

Clearly, the above reasoning also identifies a main physical factor
contributing to the fluctuations in $D$: the ever-changing plage fields in the
neighbourhood of the spot. Indeed, within the framework of the erosion model
this is the {\it only\/} possible reason for deviations from the mean law. In
reality, other contributions may originate in departures from the conditions of
the erosion model (i.e.\ from depth-independence and axisymmetry). However, as 
the erosion model proved to explain the mean decay law quite
satisfactorily, it is not unreasonable to expect that the environmental effect
of plage fields is at least one important factor in the origin of deviations
from the mean decay law.

\section{Decay rate fluctuations and the plage field}
In the previous section we argued that one of the main physical factors
contributing to the deviations from the mean decay law (\ref{eq:declaw}) is
the  environmental effect of plage fields in the neighbourhood of the spot. If
this is the case then one should expect some correlation between the daily
decay rate fluctuations and the plage flux surrounding the spot. In order to
check this prediction, we determined the mean density $B_{\mbox{\scri pl}}$ of
magnetic flux in an annulus of inner and outer radii $r_i$ and
$r_o$ around the spot from Kitt Peak magnetograms for each spot observation in
our DPR data set (when a magnetogram made on the same day was
available). 

As the magnetograms yield the line-of-sight field component only, this was
deprojected into a vertical field by dividing it by $\cos\alpha$; $\alpha$
is the heliocentric angular distance from disk center. This method clearly
introduces some extra scatter in the data but it cannot lead to systematic
errors. The polarity is conventionally taken to be ``positive'' if identical to
that of the spot in question. The results show little sensitivity to the
precise values of $r_i$ and $r_o$ chosen; here we present results calculated
with $r_i=1.5\,r_0$ and $r_o=3\,r_0$. 

\begin{figure}
%\vspace{1 cm}  % Amount of vertical space needed
\centerline{\psfig{figure=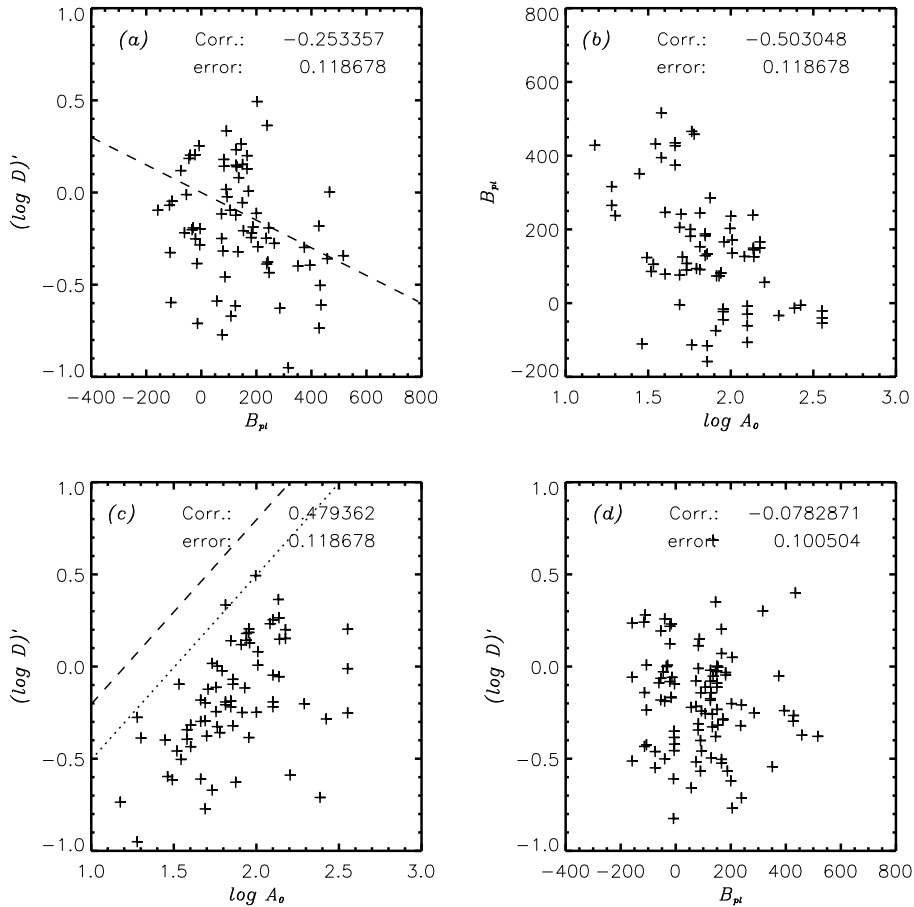,width=12.6 cm}}
\caption{Scatterplots of the mean plage field strength $B_{pl}$, in Gauss, 
around a spot
vs. $(\log D)'$ and $A_0$, in MSH, for all measurements for spots that died on the
visible hemisphere, excluding strongly non-axisymmetric spots  {\it (a, b, c),}
and for a typical simulated control data set, $a_1=a_2=0$, $a_3=0.3$  {\it
(d)}. Each panel is labeled with the value of the corresponding correlation
coefficient. The fiducial line in {\it (a)} corresponds to the dashed line in
Fig.~6 if $B_e=400\,$G. The dashed and dotted lines in {\it (c)} correspond to
spots with $(\log D)'=$const. and lifetimes of $1$ and $2$ days, respectively.}
\label{fig:plagecorr} 
\end{figure}

Figure~\ref{fig:plagecorr}a confirms our prediction of a correlation between
$B_{\mbox{\scri pl}}$ and $(\log D)'$. Indeed, the correlation coefficient is
seen to differ from zero at a confidence level well exceeding
$2\sigma$.\footnote{The standard deviation of the correlation  coefficient was
here determined with the standard formula $\sigma=N^{-1/2}$, valid in the limit
$N\rightarrow\infty$, so its value should be considered approximate. The
alternative possibility of determining the significance level by a
$t$-statistic was discarded as this method is only valid if the distribution of
the data is binormal, possibly leading to gross errors otherwise. Given our
ignorance regarding the origin and distribution of residual scatter in the
$B_{\mbox{\scri pl}}$--$(\log D)'$ relation, the use of the approximate but
robust $N^{-1/2}$ method seems safer, especially in the light of the recent
controversion concerning the significance of correlations between time serii
in the context of the solar neutrino problem (\citeNP{Oakley+}; 
\citeNP{Walther}).} \ Note that in this plot we have only included spots where
the surrounding plage field was not too far from axisymmetric. The numerical
criterion here was that, splitting the annulus in five sectors, and calculating
the mean field $B_i$ in each sector, $\ov{B_i^2}^{1/2}/\ov{B_i}<1.2$, where the
bar indicates averaging over sectors.  This selection actually only led to a
slight improvement in the correlation which was quite similar for the complete
sample.

\begin{figure}
%\vspace{1 cm}  % Amount of vertical space needed
\centerline{\psfig{figure=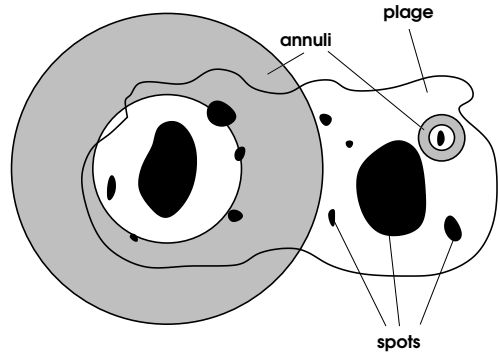,width=6.5 cm}}
\caption{Sketch illustrating the origin of the correlation in Figure~3b. The
annulus around small spots lies in the plage in its full extent, so these spots
are surrounded by the stronger plage fields.}
\label{fig:plagesketch}
\end{figure}

A doubt arises, however, concerning the origin of this correlation. It is
apparent from Figure~\ref{fig:plagecorr}b that an even stronger correlation
exists between $B_{\mbox{\scri pl}}$ and $A_0$. This correlation can be
explained by a trivial effect, illustrated in Figure~\ref{fig:plagesketch}: as
the radius of the annulus is proportional to $r_0$, for larger spots a large
part of the annulus will lie outside the plage where mean fields are weaker,
while for smaller spots the whole  annulus will fall into the strong plage. On
the other hand, owing to the observational selection effect treated in
Section~3.3.2 of Paper I (spots with higher decay rates live shorter,  and thus
will be observed on fewer occasions), another correlation appears between $\log
A_0$ and $(\log D)'$ (Fig.~\ref{fig:plagecorr}c). The coupled effect of these
two latter correlations may then be expected to give rise to a more indirect
third correlation  qualitatively similar to that seen in panel $a$. Thus the
question arises if the anticorrelation seen between the plage field and the
decay rate fluctuations is just the reflection of an observational bias, or it
has  some physical content?

To resolve this dilemma, we generated artificial data sets with the same values
of $A_0$ and $B_{\mbox{\scri pl}}$ as the real data, but with $(\log D)'$ values
generated artificially by the formula
\[ (\log D)'=a_1-a_2 B_{\mbox{\scri pl}} + a_3 x \]
where $x$ is a random variable of zero mean with a $\sigma=1$ Gaussian
probability density. In order to test our null hypothesis we first set $a_2=0$,
i.e.\ assume that no real physical correlation exists between $(\log D)'$ and
$B_{\mbox{\scri pl}}$. Then we submit our data set to a selection designed to
mimic the observational bias affecting the real data. More specifically, we
take the expected number of data for a given spot equal to its lifetime (as
observations are normally made daily), which is in turn calculated assuming
that $(\log D)'$ stays constant during the decay of the spot. 
%(Note that this latter assumption implies a
%stronger bias than in reality, as a spot with a higher-than-average
%decay rate is in fact likely to show a less pronounced decay rate on other days,
%increasing its lifetime and observation probability. On the other hand, the fact
%that our method implies some finite observational probability for spots of any
%short lifetime, whereas in reality obviously at least 2 consecutive observations
%are needed to determine a decay rate, approximately compensates for the first 
%mentioned effect.) 

The parameters $a_1$ and $a_3$ are freely adjustable, but subject to the
condition that the resulting range of ``observed'' $(\log D)'$ values should
coincide with the real range. After having computed a large number of
simultaneous distributions with a wide variety of $a_1$ and $a_3$ values we had
to conclude that no parameter combination is able to yield a correlation
coefficient higher than about $0.1$ in typical realizations. The optimal
agreement was found in the case $a_1=a_2=0$, $a_3=0.3$, shown in
Figure~\ref{fig:plagecorr}d.  It is apparent that the correlation coefficient
does not differ from zero by more than $1\sigma$. It seems therefore that the
indirect effect of the correlations shown in panels $b$ and $c$ is insufficient
to fully explain the correlation found in panel $a$, part of which must be a
real physical effect. Indeed, if in our simulated data we allow $a_2\neq 0$ it
becomes straightforward to find a good representation of the data: $a_1=0$,
$a_2=0.3$, $a_3=0.4$ is for instance just such a case (not shown here). 

We thus conclude that while selection effects probably contribute to the
anticorrelation of $(\log D)'$ with the plage field strength around spots, a
part of this effect is likely to have a physical origin.

\section{Theoretical interpretation}
In order to understand the origin of a physical anticorrelation between $(\log
D)'$ and the value of the external magnetic field strength
%, the existence of which is indicated by our studies in previous section, 
here we extend the turbulent erosion model of sunspot decay for the case when
the flux tube is embedded in a preexisting parallel homogeneous magnetic field.
Apart from this generalization, we consider the same cylindrically symmetric
case as in PM97, i.e.\ we solve the nonlinear diffusion equation
\begin{equation} 
  \frac{\partial}{\partial t}\left(\rvar B\right)=\frac{\partial}{\partial\rvar}
  \left[\rvar\;\nu(B)\;\frac{\partial B}{\partial\rvar}\right]
  \label{eq:diff} 
\end{equation}
where $B$ is the 
magnetic flux density, $\rvar$ is the radial coordinate, and $\nu$ is the 
turbulent magnetic diffusivity. For the dependence of the diffusivity on the 
magnetic field we still consider the same function satisfying the basic physical 
requirements:
\begin{equation} 
  \nu(B)=\frac{\nu_0}{1+|B/B_{\mbox{\scriptsize e}}|^{\alpha_{\nu}}} 
  \label{eq:nuexpr} . 
\end{equation}
Here $\nu_0$ is the unperturbed value of the diffusivity, and $\alpha_\nu$ is 
a parameter quantifying the steepness of the diffusivity 
cutoff near $B_{\mbox{\scriptsize e}}$, the latter being the field strength 
where the diffusivity is reduced by 50\,\%. Physically, one expects 
$B_{\mbox{\scriptsize e}}\sim B_{\mbox{\scriptsize eq}}$, 
the kinetic energy density of photospheric turbulence. 

Using the notation $b\equiv B/B_{\mbox{\scriptsize e}}$, our initial conditions 
will now be
\begin{equation} 
 b(r)=\bext+\frac{b_0-\bext}{1+(r/r_0)^{22}}  
 \label{eq:B0expr1} 
\end{equation} 
where $b_0$ is the normalized field strength at the center of the tube, and 
$\bext$ is the external (``plage'') field strength: we now assume  $\bext\neq
0$. In this formula $r_0$ is the initial radius of the tube at time $t=0$; for
a comparison with observations it may be identified with the maximal equivalent
radius of the spot.

\begin{figure}
 \centerline{\psfig{figure=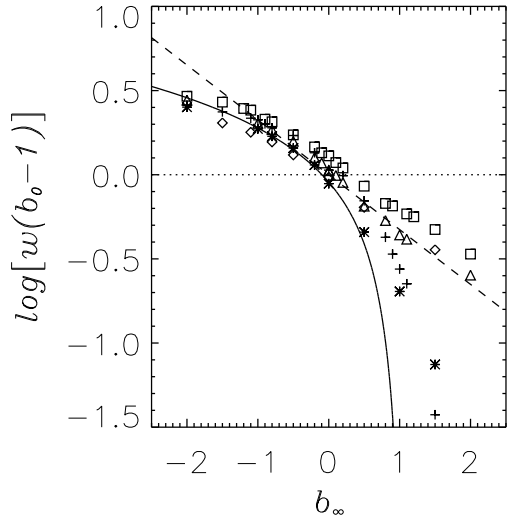,width=8.0 cm}}
 \caption{Current sheet velocity $w$ as a function of the background
 field strength $\bext$.  
 $\Diamond$: $\alpha_\nu=2.3$, $b_0=30$; $\Box$: $\alpha_\nu=2.5$, $b_0=10$;
 $\triangle$: $\alpha_\nu=3.0$, $b_0=10$; $\ast$: $\alpha_\nu=4.5$, $b_0=10$;
 $+$: $\alpha_\nu=7$, $b_0=3$. The dashed line is just a fiducial straight line,
 drawn to aid the eye. Solid line: eq. (B-3)}
 \label{fig:theorpred}
\end{figure}

The values of the current sheet velocity $w$ are plotted against $\bext$ for
different numerical solutions of Equations (\ref{eq:diff})--(\ref{eq:B0expr1}) 
in Figure~\ref{fig:theorpred}. It is apparent that for
$\bext=0$ all the solutions are in agreement with the formula 
\begin{equation} 
   w\simeq  C {\nu_0}/{r_0}(b_0-1) ,  \label{eq:wexpr} 
\end{equation}
with $C\simeq 1.2$, as expected. 

With intermediate values of $\alphanu\sim 2$--$3$ the dependence of $w$ 
on $\bext$ is approximately logarithmic for $\bext\ga -1$, while for 
$\alphanu >4$
it is strongly reduced for $\bext >0$. The high-$\alphanu$ behaviour
may be interpreted in terms of a generalization of the dimensional
argument leading to (\ref{eq:wexpr}), see Appendix B. 

The actual value of $\alphanu$ is poorly known, but from
Figure~\ref{fig:theorpred} it is clear that in the range $-1<\bext<1$ the rate
of flux loss should depend  approximately logarithmically on $\bext$,
independently of the value of $\alphanu$. In other words, the background field
acts as multiplicative noise in the decay process. Setting 
$B_{\mbox{\scriptsize e}}\simeq B_{\mbox{\scriptsize eq}}\simeq 400$G,  the
photospheric value, this logarithmic relation is plotted over the observational
data in Figure~\ref{fig:plagecorr}a. The apparent good agreement lends further
support to the physical reality of the correlation found in the previous
section.

\section{Conclusion}

\subsection{Proposition: decay rate deviations as effects of the plage field}
The theoretical results presented in Section~5 suggest the following scheme for
the origin of random deviations from the mean parabolic decay law
(\ref{eq:declaw}). It is rather natural to assume that the value of the
background plage field strength of independent origin shows a  Gaussian
probability distribution around some expected value around a spot.
Figure~\ref{fig:theorpred} then implies that the flux loss rate from the  flux
tube should be characterized by a lognormal distribution (just as we found in
Section~2), and that an anticorrelation should exist between the logarithmic
decay rate fluctuations $(\log D)'$ and the strength of the plage field
surrounding the spots (just as we found in Section~4). The generalization of
the erosion model to the case of a tube embedded in an external homogeneous
magnetic field then apparently offers a simple and elegant explanation for the
presence of large lognormally distributed deviations from the mean decay rate
in individual spots.

It should be noted that a relation between the decay rate of sunspots and their
environment was also suggested by \citeN{Antalova+Macura}. While
\citeN{FMI+MVA} correctly pointed out that the existence of a bimodal
distribution of decay rates (i.e. well-separated branches of fast- and 
slow-decaying spots in the area--decay rate plane) in that study was the
consequence of  a trivial selection effect, the apparent physical difference
found by \citeN{Antalova+Macura} between the two classes (slow-decaying spots
tending to be ``naked'' spots in the sense defined by \citeNP{Liggett+Zirin})
still remains, and it may possibly be related to the environmental effect
proposed in the present work.

\subsection{Caveats and call for more data} However attractive the scheme
outlined in the previous subsection may seem, it is almost certainly an
oversimplification of the real situation. Firstly, it is obvious from
Figure~\ref{fig:plagecorr}a that a large part of the scatter in $(\log D)'$ is
independent of $B_{\mbox{\scri pl}}$. Though part of this scatter may be due to
our method of deriving the vertical field component by deprojecting the
line-of-sight component measured, etc., one may expect that effects like
deviations from axial symmetry and sub-photospheric forces should necessarily
contribute a great deal to deviations from the relation predicted by the
erosional model. It is therefore certainly not the case that the lognormal
distribution of decay rates can be explained by the plage effect {\it alone:}
however, owing to the central value theorem of probability theory, the
lognormal distribution can be saved if the other influences also act as
multiplicative noise.

Another caveat is related to the strong influence of the bias treated in
Section~3.3.2 of Paper I throughout this analysis. Though, as we have seen, it
is  not impossible to correct for these effects, some fine points and
simplifications in the correction process may always leave some doubt
concerning the robustness of the conclusions derived. The main difficulties
here are all due to the fact that our data (like all solar patrols realized in
this outgoing century) were taken on a daily basis. One of the most emphatic
conclusions of this study is that \it observations made on a daily basis are
insufficient for an in-depth study of the sunspot decay process. \/ \rm What is
needed is at least one observational campaign lasting several months,
preferably at the time of solar maximum, with several {\it simultaneous\/}
white-light heliogram {\it and\/}  magnetogram observations a day. Only from
the use of such higher quality data can we expect the final illumination of
the sunspot decay problem.

\acknowledgements 
This work was funded by the DGES project no.~95-0028, by the OTKA under grant 
no.\ T17325, and by the FKFP project no 0201/97. 
The use of public domain 
NSO Digital Library data (Kitt Peak magnetograms) is acknowledged.

\begin{center}
\bf Appendix A: A comparison of DPR and Mt Wilson sunspot areas
\end{center}

\begin{figure}[ht]
 \centerline{\psfig{figure=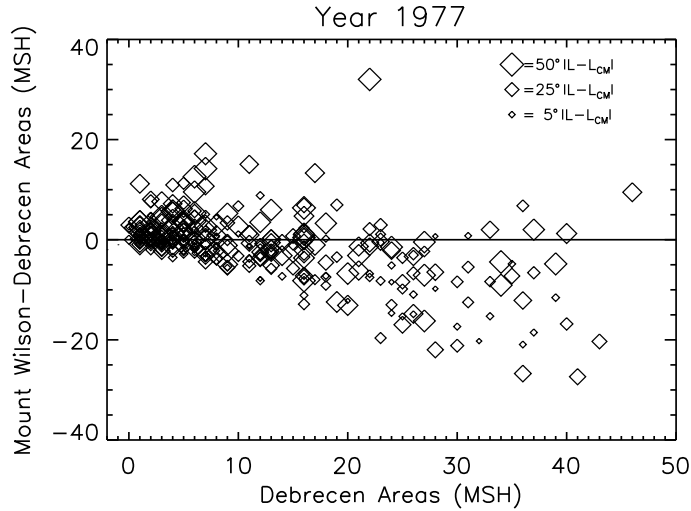,width=10 cm}}
 \caption{Total spot areas from the DPR 1977 plotted against their difference
 from the areas calculated by multiplying by 6.2 the corresponding umbral areas 
 of the Mt Wilson catalogue. The size of the symbols is related to the central
 meridian (CM) distance of the spot, $|L-L_{CM}|$; the scale of symbol sizes is
 set by three examples given in the top right corner.}
 \label{fig:DPRWilson}
\end{figure}

In an attempt to extend our database we examined the possibility of
incorporating Mt Wilson observations in our data (\citeNP{Howard+:MtWilson}). 
One problem to be solved here was that in the Mt Wilson catalogue the spots are
not identified from one day to the next, so in order to calculate decay rates
the identification had to be done. We developed an algorithm for the
identification of spots taking into account the differential rotation. A
comparison of the results with the published DPR data, where available, showed
that the identification has an acceptable rate of success, at least for the
larger spots in each active region (which tend to dominate our data set). 

A further difficulty is caused however by the fact that the Mt Wilson data only
yield estimated umbral areas. Umbral areas are notoriously more difficult to
determine than umbra+penumbra areas (\citeNP{Gyori:umbrarea}); besides, the
method of determination applied was rather approximate
(\citeNP{Howard+:MtWilson}). While it is possible to  translate umbral areas
$A_U$ into total areas $A$ by the simple linear rule  $A=6.2\,A_U$ (the value
of the coefficient was determined from an analysis of DPR area data),  this
linear relation involves a large scatter, further contributing to the
uncertainty of the area values. The resulting large errors are borne out in
Figure~\ref{fig:DPRWilson}, where we plotted the difference between the
MtWilson area values thus derived and the corresponding  DPR area values
against DPR areas for 1977. It is apparent that the scatter is very large
indeed, effectively excluding the use of Mt Wilson data for our present
purpose. 

\begin{figure}[ht]
 \centerline{\psfig{figure=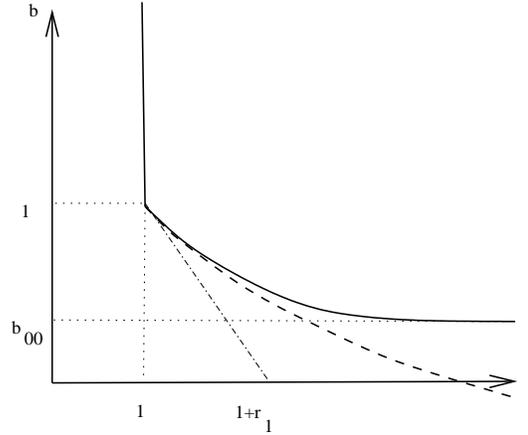,width=7.0 cm}}
 \caption{Sketch showing the characteristic shape and scales of the
 magnetic field profile (solid) and of the stationary solution (B-1) (dashed)
 near the outer edge of the current sheet. $r_1\equiv b/(db/dr)$ at $r=1$}
 \label{fig:analytic}
\end{figure}

It may be worth noting that Figure~\ref{fig:DPRWilson} also seems to suggest a 
systematic trend: the areas computed from Mt Wilson data are lower than their
DPR counterparts for large spots, while the reverse tends to be the case with
small spots. One possible interpretation could be that the fixed ratio
$A/A_U=6.2$ used  here may be incorrect, and that the umbral/total area ratio
may in fact have some mean area dependence. However, a similar plot constructed
using the DPR umbral areas instead of the Mt Wilson areas does not show any
obvious trend (while it still shows a high scatter), which seems to exclude
this explanation. The trend mentioned is then probably due to 
differences in the applied methods of area measurement.

\begin{center}
\bf Appendix B: Analytical interpretation of the high-$\alpha_\nu$ behaviour 
of the generalized erosion models
\end{center}

\renewcommand{\theequation}{B-\arabic{equation}}
\setcounter{equation}{0}

For simplicity we set the units of time and length so that $\nu_0=1$ and
$b(1)=1$. Restricting our interest to the case $b_0\gg 1$, from
(\ref{eq:wexpr}) $w\ll 1$ follows, so that in the range $1<r<1+r_1$ the
solution may be approximated as the stationary solution of the linear diffusion
equation in cylindrical geometry (cf.\ Fig.~\ref{fig:analytic}): 
\begin{equation} 
  b=1-\frac{\ln r}{r_1}  \label{eq:logprofile}  
\end{equation}
From the numerical solutions we know that the erosional (i.e.\ nearly
form-invariant propagating) solution forms after an initial transition time
$\ti\sim 1$. If the initial profile at $t=0$ is  sufficiently steep  [as is the
case with (\ref{eq:B0expr1})] then at $t=\ti$ the amount of magnetic flux
outside $r=1$ equals the flux lost from within the CS, and the original plage 
flux there: 
\begin{equation} 2\pi\int_1^{1+r_1}rb(r)\,dr\simeq \pi b_0(r_0^2-1) +\pi 
  \bext(r_1^2+2r_1) 
\end{equation}
As we are still within about one diffusive timescale from
$t=0$, practically all this flux must still lie inside the radius $1+r_1$,
where  the approximation (\ref{eq:logprofile}) may be used for $b(r)$.
On the other hand, $r_0=1+w\ti$, while $w$ and $r_1$ are related by 
$w\simeq [r_1(b_0-1)]^{-1}$ (PM97, eq.\ (10)), so,
neglecting a term of order $b_0^{-2}$, we have
\begin{equation} (r_1^2+2r_1)\left(1+\frac 1{2r_1}-\bext\right) 
   -\frac{(1+r_1)^2}{r_1}\ln (1+r_1)-\frac{2b_0}{b_0-1}r_1=0
   \label{eq:r1}
\end{equation}
Setting $\ti=1$ and $b_0=10$, Equation (\ref{eq:r1}) can be 
solved numerically
for $r_1$, and thus for $w\simeq [r_1(b_0-1)]^{-1}$; the solution
turns out to be a very nearly linear function of $\bext$ (solid curve in 
Figure~\ref{fig:theorpred}).

%\section{dg}

%\bibliography{krisaj,kriskp,krisrz}

\begin{thebibliography}{}

\bibitem[\protect\citeauthoryear{{Antalov\'a} and {Ma\v
  cura}}{1985}]{Antalova+Macura}
{Antalov\'a}, A. and {Ma\v cura}, R.: 1985,
\newblock {\em {Contr.\ Skaln.\ Pleso Obs.}\/} {\bf 14}, 163

\bibitem[\protect\citeauthoryear{{Dezs{\H o}}, Gerlei, and
  {Kov\'acs}}{1987}]{DPR}
{Dezs{\H o}}, L., Gerlei, O., and {Kov\'acs}, {\'A}.: 1987,
\newblock {\em Debrecen Photoheliographic Results for the year 1977\/},
\newblock Publ.\ Debrecen Heliophys.\ Obs., Heliogr.\ Series No.~1, Debrecen

\bibitem[\protect\citeauthoryear{{Dezs{\H o}}, Gerlei, and
  {Kov\'acs}}{1996}]{DPR78}
{Dezs{\H o}}, L., Gerlei, O., and {Kov\'acs}, {\'A}.: 1996,
\newblock {\em Debrecen Photoheliographic Results for the year 1978\/},
\newblock Publ.\ Debrecen Heliophys.\ Obs., Heliogr.\ Series No.~2,
  ftp://fenyi.sci.klte.hu/pub/DPR/1978

\bibitem[\protect\citeauthoryear{Gy{\H o}ri}{1998}]{Gyori:umbrarea}
Gy{\H o}ri, L.: 1998,
\newblock {\em \solphys\/} {\bf 180}, 109

\bibitem[\protect\citeauthoryear{Howard}{1992}]{Howard:decay}
Howard, R.: 1992,
\newblock {\em \solphys\/} {\bf 137}, 51

\bibitem[\protect\citeauthoryear{Howard, Gilman, and
  Gilman}{1984}]{Howard+:MtWilson}
Howard, R., Gilman, P.~A., and Gilman, P.~I.: 1984,
\newblock {\em \apj\/} {\bf 283}, 373

\bibitem[\protect\citeauthoryear{Liggett and Zirin}{1983}]{Liggett+Zirin}
Liggett, M. and Zirin, H.: 1983,
\newblock {\em {Solar Phys.}\/} {\bf 84}, 3

\bibitem[\protect\citeauthoryear{{Mart{\'\i}nez Pillet}, {Moreno-Insertis}, and
  V\'azquez}{1993}]{VMP+:periph.decay}
{Mart{\'\i}nez Pillet}, V., {Moreno-Insertis}, F., and V\'azquez, M.: 1993,
\newblock {\em \aap\/} {\bf 274}, 521 (MMV93)

\bibitem[\protect\citeauthoryear{{Moreno-Insertis} and
  {V{\'a}zquez}}{1988}]{FMI+MVA}
{Moreno-Insertis}, F. and {V{\'a}zquez}, M.: 1988,
\newblock {\em \aap\/} {\bf 205}, 289

\bibitem[\protect\citeauthoryear{Oakley {\em et~al.}}{1994}]{Oakley+}
Oakley, D.~S., Snodgrass, H.~B., Ulrich, R.~K., and VanDeKop, T.~L.: 1994,
\newblock {\em \apj\/} {\bf 437}, L63

\bibitem[\protect\citeauthoryear{Petrovay}{1998}]{Petrovay:supdif}
Petrovay, K.: 1998,
\newblock A crossroads for european solar and heliospheric physics: recent
  achievements and future mission possibilities, ESA Publ.\ SP-417, p.~273

\bibitem[\protect\citeauthoryear{Petrovay and
  {Moreno-Insertis}}{1997}]{Petrovay+FMI:erosion}
Petrovay, K. and {Moreno-Insertis}, F.: 1997,
\newblock {\em \apj\/} {\bf 485}, 398 (PM97)

\bibitem[\protect\citeauthoryear{Petrovay and {van
  Driel-Gesztelyi}}{1997}]{Petrovay+vDG:decay1}
Petrovay, K. and {van Driel-Gesztelyi}, L.: 1997,
\newblock {\em {Solar Phys.}\/} {\bf 176}, 249 (Paper I)

\bibitem[\protect\citeauthoryear{Walther}{1998}]{Walther}
Walther, G.: 1998,
\newblock {\em \apj\/} {\bf 513}, 990

\end{thebibliography}
%\bibliographystyle{solphys}

\begin{ao}
\\
K. Petrovay\\
Instituto de Astrof\'\i sica de Canarias\\
La Laguna, Tenerife, E-38200 Spain\\
E-mail: kpetro@iac.es
\end{ao}

\end{article}
\end{document}